\documentclass[12pt]{article}%
\usepackage{amsmath,amsfonts,amssymb,latexsym,graphicx}
\usepackage{caption}
\usepackage{subcaption}

\providecommand{\U}[1]{\protect\rule{.1in}{.1in}}

\def\be{\begin{equation}}
\def\ee{\end{equation}}
\def\bq{\begin{eqnarray}}
\def\eq{\end{eqnarray}}
\def\beq{\begin{eqnarray}}
\def\eeq{\end{eqnarray}}

\begin{document}

\title{\textsc{The cosmological frame principle and cosmic acceleration}}
\author{{\Large {\textsc{Spiros Cotsakis$^{1,2}$}\thanks{\texttt{skot@aegean.gr}},
\textsc{Jose P. Mimoso}$^{3}$\thanks{\texttt{jpmimoso@fc.ul.pt}}},
\textsc{John Miritzis}$^{4}$\thanks{\texttt{imyr@aegean.gr}}}\\
$^{1}$Institute of Gravitation and Cosmology, RUDN University\\ul. Miklukho-Maklaya 6, Moscow 117198, Russia\\
$^{2}$Research Laboratory of Geometry, Dynamical Systems\\and Cosmology, University of the Aegean,\\Karlovassi 83200, Samos, Greece\\
$^{3}$Departamento de F\'{\i}sica and Instituto de Astrof\'{\i}sica e Ci\^encias do Espa\c co\\
Faculdade de Ci\^encias, Universidade de Lisboa\\
Ed. C8, Campo Grande, 1769-016 Lisboa, Portugal\\
$^{4}$Department of Marine Sciences\\University of the Aegean\\University Hill, Mytilene 81100, Greece}
\date{August 2023}
\maketitle

\begin{abstract}
\noindent We discuss  implications of the cosmological frame principle which
states that cosmological effects of modified gravity must be stable as solutions of each of the corresponding sets of dynamical equations holding in the two conformally-related frames. We show that there are such globally stable,
`frame-independent' solutions describing cosmic acceleration, suggesting that they may
represent a physically relevant effect. This result highlights the importance
of further investigation into the implications of the  frame
principle for cosmological properties that rely on the use of conformal frames.

\end{abstract}

\section{Introduction}

Modified gravity has become an essential part of theoretical cosmology as an
extension of general relativity to understand the nature of gravitational
effects in the very early and very late universe, see e.g., \cite{cal} -
\cite{bbccft} for recent reviews. The conformal potential approach to modified
gravity casts the theory in an alternative form by performing a conformal
transformation of the original field equations in the Jordan frame to the
conformally related Einstein frame representation of the theory as general
relativity plus a self-interacting scalar field \cite{cot22,cy22}. This was first
introduced for the Brans-Dicke theory \cite{d1} by Dicke in \cite{d2}, for the
$f(R)$ gravity theory of \cite{ba-ot} in \cite{B-C88}, and for the
scalar-tensor extension in \cite{ma89}; for recent investigations cf.
\cite{pali}-\cite{baco}. 

As is very common in the literature of this vast subject, using the conformal frames one may work at will or
choice in any of the two conformally-related theories. In particular, one suspects that even if cosmological phenomena may look different in the two conformally related frames, yet relations between observables must be the same (of course, a mathematical equivalence does not imply a physical one), cf. \cite{k1}-\cite{k3} and related refs. therein. In these references, a procedure of corresponding effects in the two frames is successfully applied to a variety of deep cosmological questions, such as the singularity problem,  the isotropization issue in anisotropic models, or the problem of matching solutions before and after a singularity `crossing'. These interesting analyses suggest that one may use the two frames in a productive way to treat cosmological problems through field reparametrization in a similar way as when  we have different coordinates systems in classical mechanics. These analyses imply that conformal frames are more than just a mathematical procedure, and possibly suggest the existence of some yet-unknown underlying physical effect taking place between the two conformal frames.

In this paper, we suggest a more precise formulation of this effect which we coin the cosmological frame principle, with the following formulation: 

The cosmological frame principle: \emph{All cosmological solutions of modified gravity are
frame-independent, that is they should preserve their stability properties as solutions of the two conformally-related sets of dynamical equations in the two frames.}

The notion of stability in this statement requires some discussion. We mean stability in any given well-defined,  mathematical sense, that is Liapunov, asymptotic, orbital, asymptotic orbital, or structural stability of the two systems (or its generalization to system families). Under the frame principle, we propose to accept as a physical effect one that proves stable in \emph{any} one of the above ways.
Therefore a way to test the frame principle in concrete situations is to check the
stability of solutions describing the same property in both frames (in the case of structural stability of families, testing the frame principle would imply a full bifurcation analysis of the two systems). If a
particular property is proven to be stable as a solution of the equations in some way in both frames (that is for both systems of conformally-related dynamical equations), then it can be regarded
as \emph{frame-independent} in the sense that it exists independently from which
frame one uses to study it, despite if it appears differently in the two frames. (It is important to emphasize the use of stable \emph{solutions} of the two sets of conformally-related dynamical equations in the statement of the frame principle, not just arbitrary functions connected by the conformal transformation between the two frames.)

A prime example is inflation in various contexts
(cf. \cite{star1} and also \cite{B-C88,ma89,k1}), or various cosmic no-hair
theorems which give conditions under which de Sitter or quasi-de Sitter spaces
are stable in both the $f(R)$ and the Einstein frames, cf.
\cite{ba-ot,star,mms,cf-93,cm98}, and also the stability of more general
Friedmann--Lema\^{\i}tre--Robertson--Walker (FLRW) solutions in both frames
\cite{ba-ot,cf-93b}. Another example is the viability of an $f\left(
R\right)  $ theory, for instance, $f(R)=R-\mu^{2(n+1)}/R^{n}$, with
$\mu>0,n>1$, promoted to explain the late-time cosmic acceleration
\cite{ccct,cdtt}, for which in the Einstein frame for FLRW models, it was shown in
\cite{tzmi} that in the limit $n\rightarrow\infty$, these models are not
cosmologically viable. This fact reinforces a general feeling that models with
$f(R)=R-\mathrm{const}/R^{n}$ are cosmologically unacceptable \cite{agpt}. We note that it would be an interesting result if one could show the stability of the matching solutions found in \cite{k2,k3} and related references therein, that is show the validity of the cosmological frame principle in the singularity crossing problem.

In this paper, we provide a further example of the use of the frame principle
in cosmology, namely, the independence of the property of future acceleration on the
choice of frame. We study cosmic acceleration using the frame principle as a
guide, and stability analysis in both the Jordan and Einstein frames for flat
FLRW universes in the setup of a quadratic Lagrangian gravity theory with a
cosmological constant. To this end, in the following two sections we
investigate the evolution of FLRW flat models in the quadratic theory with a
non-zero cosmological constant of the form $R+\epsilon R^{2}-2\Lambda$ firstly
in the Jordan frame (next section) and then in the Einstein frame (Section 3).
We prove rigorously that acceleration is a stable property in both frames, and
comment on the validity of recollapsing universes with a negative cosmological
constant independently of the frame chosen. For simplicity we restrict our analysis to vacuum models, and leave for subsequent work the inclusion of a perfect fluid given its further mathematical intricacies.  Our stability proof in this paper reinforces the view that future
acceleration is a typical property for these models, and it  adds to the physical
plausibility of dark energy.

\section{Acceleration in the Jordan frame}

We adopt the metric and curvature conventions of \cite{wael}, and choose units
so that $c=1=8\pi G$. For the quadratic theory, $R+\epsilon R^{2}-2\Lambda$ in
a flat FLRW model, the $00$ equation is,
\begin{equation}
H^{2}+2\epsilon\left(  RH^{2}+H\dot{R}-\frac{R^{2}}{12}\right)  =\frac
{\Lambda}{3}, \label{00}%
\end{equation}
while the evolution equation for $H$ takes the form,%
\begin{equation}
\dot{H}=\frac{1}{6}R-2H^{2}. \label{hubb}%
\end{equation}
We shall also make use of the trace equation which reads,
\begin{equation}
\ddot{R}+3H\dot{R}+\frac{1}{6\epsilon}R=\frac{2\Lambda}{3\epsilon}.
\label{trac}%
\end{equation}
With the variables $R\ $and $\dot{R}$, equations (\ref{trac}) and (\ref{hubb})
constitute a three-dimensional system. In the following we shall assume that $\epsilon>0$. The non-zero constant $\epsilon$ may be
used to define dimensionless variables by the rescaling,%
\begin{equation}
R=\frac{2}{\epsilon}x_{1},\ \ \dot{R}=\sqrt{\frac{2}{3\epsilon^{3}}}%
~x_{2},\ \ H=\frac{h}{\sqrt{6\epsilon}},\ \ t=\sqrt{6\epsilon}~\tau,
\end{equation}
and our system becomes,%
\begin{align}
\dot{x}_{1}  &  =x_{2},\nonumber\\
\dot{x}_{2}  &  =-x_{1}-3x_{2}h+\frac{\lambda}{4},\label{jord}\\
\dot{h}  &  =2x_{1}-2h^{2},\nonumber
\end{align}
where the dot denotes differentiation with respect to $\tau$ and
$\lambda=8\epsilon\Lambda$. The constraint (\ref{00}) takes the form,
\begin{equation}
h^{2}+4x_{1}h^{2}+4hx_{2}-4x_{1}^{2}=\lambda/4. \label{const2}%
\end{equation}
The only feasible equilibrium point of (\ref{jord}) is
\begin{equation}
\mathbf{q}=\left(  x_{1}=\lambda/4,x_{2}=0,h=\sqrt{\lambda}/2\right)  ,
\label{eq1}%
\end{equation}
and corresponds to a late accelerating universe.

\begin{figure}[tbh]
\begin{center}
\includegraphics{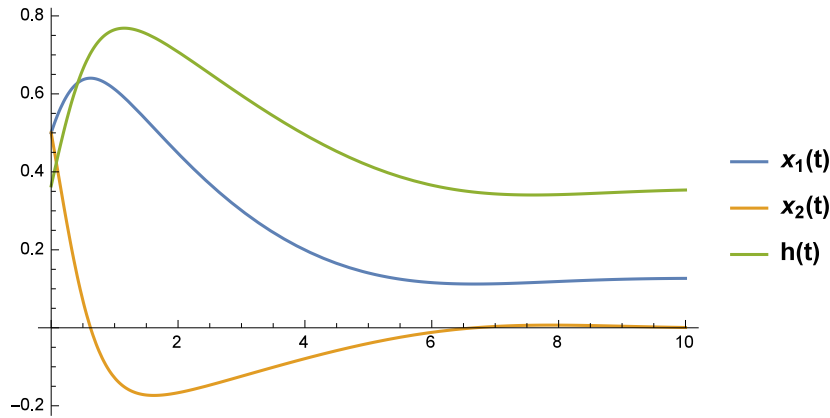}
\end{center}
\par
\caption{Solutions of (\ref{jord}) for $\lambda=1/2$. The initial values are
chosen so that they respect the constraint (\ref{const2}).}%
\label{solu2}%
\end{figure}

It turns out that all of the eigenvalues of the Jacobian matrix of
(\ref{jord}) at $\mathbf{q}$ have negative real parts, therefore $\mathbf{q}$
is locally asymptotically stable and attracts all nearby solutions. In Figure
\ref{solu2}, the solutions $x_{1}\left(  t\right)  $, $x_{2}\left(  t\right)
$ and $h\left(  t\right)  $, are shown approaching their limiting values
$1/8$, $0$ and $\sqrt{1/2}/2$ respectively.

%

\section{Acceleration in the Einstein frame}

We now consider the quadratic theory $R+\epsilon R^{2}-2\Lambda$ in the
conformal frame. The contents of this section follow the path of ideas in
\cite{miri3}, but for the sake of completeness we give a brief presentation.
The evolution of flat FLRW models is described by the Friedmann equation,
\begin{equation}
H^{2}=\frac{1}{3}\left(  \frac{1}{2}\dot{\phi}^{2}+V\left(  \phi\right)
\right)  , \label{fri1jm}%
\end{equation}
the Raychaudhuri equation,
\begin{equation}
\dot{H}=-H^{2}-\frac{1}{3}\dot{\phi}^{2}+\frac{V}{3}, \label{fri2jm}%
\end{equation}
and the equation of motion of the scalar field,
\begin{equation}
\ddot{\phi}+3H\dot{\phi}+V^{\prime}\left(  \phi\right)  =0. \label{emsjm}%
\end{equation}
Here $V\left(  \phi\right)  $ is the potential energy of the scalar field
associated with the conformal transformation and $V^{\prime}=dV/d\phi.$ We
assume that the universe is initially expanding, i.e. $H\left(  0\right)  >0$.
Then one can show by standard arguments, \cite{fost,miri}, that the universe
remains ever-expanding, i.e. $H\left(  t\right)  \geq0$ for all $t\geq0$.
Using the constraint equation (\ref{fri1jm}), the evolution equation for $H$
simplifies to%
\begin{equation}
\dot{H}=-\frac{1}{2}\dot{\phi}^{2}. \label{frie2}%
\end{equation}
Equation (\ref{frie2}) implies that $H$ is a decreasing function of time $t$
and is bounded from below either by $0$ or $\sqrt{V(\phi_{\ast})/3})$ where
$V^{\prime}(\phi_{\ast})=0$.

\begin{figure}[tbh]
\begin{center}
\includegraphics[scale=0.7]{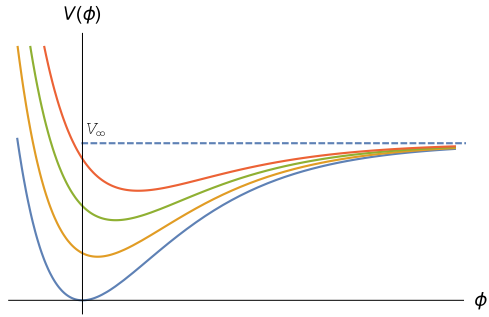}
\end{center}
\caption{The family of potentials (\ref{pote1}) for $V_{\infty}=1$ and
$\Lambda$ taking values $0,0.3,0.6,0.9$. }%
\label{fig1}%
\end{figure}

For the quadratic theory $R+\epsilon R^{2}-2\Lambda$ the corresponding family
of potentials in the Einstein frame is
\begin{equation}
V_{\Lambda}\left(  \phi\right)  =V_{\infty}\left(  1-e^{-\sqrt{2/3}\phi
}\right)  ^{2}+\Lambda e^{-2\sqrt{2/3}\phi}, \label{pote1}%
\end{equation}
with $V_{\infty}=1/8\epsilon$. For all $\Lambda>0,$ the functions $V_{\Lambda
}\left(  \phi\right)  $ have a positive minimum $V_{\min}$ at some $\phi
_{m}>0$, but otherwise share the same qualitative behaviour as if this term
were absent, see Figure \ref{fig1}.

We simplify the system by rescaling the variables as follows,%
\begin{equation}
\phi\rightarrow\sqrt{3/2}\,\phi,\ \ \dot{\phi}=\sqrt{2V_{\infty}%
}\,\,y,\ \ H=\sqrt{\frac{4V_{\infty}}{3}}\,h,\ \ t=\sqrt{\frac{3}{4V_{\infty}%
}}\,\tau. \label{resc1}%
\end{equation}
Furthermore, in order to take account of the equilibrium point corresponding
to the point at \textquotedblleft infinity\textquotedblright\ and to remove
the transcendental functions, it is convenient to introduce the variable $u$
defined by,
\begin{equation}
u:=e^{-\phi}, \label{u}%
\end{equation}
and the system (\ref{fri2jm})-(\ref{emsjm}) finally becomes,
\begin{align}
\dot{u}  &  =-uy,\nonumber\\
\dot{y}  &  =-u+\left(  1+\lambda\right)  u^{2}-3hy,\label{sys3}\\
\dot{h}  &  =-h^{2}-\frac{1}{2}y^{2}+\frac{1}{4}\left(  1-u\right)  ^{2}%
+\frac{1}{4}\lambda u^{2},\nonumber
\end{align}
where the dot denotes differentiation with respect to $\tau$ and
$\lambda=\Lambda/V_{\infty}\equiv8\epsilon\Lambda$. Note that under the
transformation (\ref{u}), the resulted three-dimensional dynamical system
(\ref{sys3}) is quadratic. The constraint (\ref{fri1jm}) takes the form%
\begin{equation}
h^{2}=\frac{1}{4}\left(  y^{2}+\left(  1-u\right)  ^{2}+\lambda u^{2}\right)
. \label{const}%
\end{equation}

In the study of the equilibrium points we note that $u=1$ corresponds to
$\phi=0$ and $u=0$ corresponds to $\phi=\infty,$ i.e. to the flat plateau of
the potential. There are two equilibrium points of (\ref{sys3}):
\[
\mathrm{EQ1}:\left(  u=0,y=0,h=\frac{1}{2}\right)  .
\]
This corresponds to the de Sitter universe with a cosmological constant equal
to $\sqrt{V_{\infty}/3}$, i.e. the scalar field stays at the flat plateau of
the potential. A necessary condition for the existence of this equilibrium, is
that the scalar field reaches the flat plateau, which is impossible if we
restrict ourselves to initial values of $H$ smaller than $\sqrt{V_{\infty}/3}%
$.
\begin{equation}
\mathrm{EQ2}:\left(  u=\frac{1}{\lambda+1},y=0,h=\frac{1}{2}\sqrt
{\frac{\lambda}{\lambda+1}}\right)  . \label{eq2}%
\end{equation}
This corresponds to a late accelerating universe while the scalar field
reaches the value $\phi_{m}=\ln\left(  1+\lambda\right)  $, corresponding to
the minimum of the potential.

EQ2 is the most interesting case because, if one could prove that it is
stable, this would imply that the late accelerating expansion solution,
attracts all nearby solutions. In fact, it is easy to see that the eigenvalues
of the Jacobian matrix of (\ref{sys3}) at EQ2 are two complex conjugate with
negative real parts and one real and negative. Therefore the equilibrium point
(\ref{eq2}) of (\ref{sys3}) is locally asymptotically stable.

\begin{figure}[tbh]
\begin{center}
\includegraphics{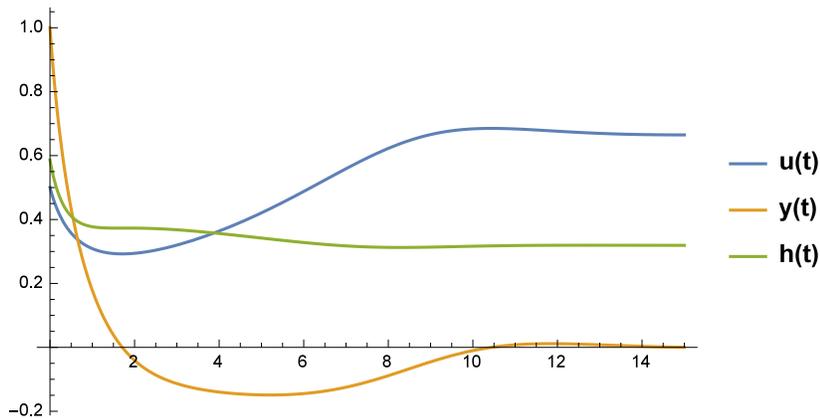}
\end{center}
\par
\caption{Solutions of (\ref{sys3}) for $\lambda=1/2$. The initial values are
chosen so that they respect the constraint (\ref{const}). Note that $u\left(
t\right)  $ and $y\left(  t\right) $ exhibit damped oscillations.}%
\label{solu}%
\end{figure}

In Figure \ref{solu}, the solutions $u\left(  t\right)  $, $y\left(  t\right)
$ and $h\left(  t\right)  $ approach their limiting values $2/3$, $0$ and
$1/\left(  2\sqrt{3}\right)  $ respectively. These results mean that near the
equilibrium EQ2, the dumped oscillations of the scalar field settle down to
its minimum value while the Hubble function achieves its constant limiting
value $\sqrt{V(\phi_{\min})/3}$. 

\section{Discussion}
We have analyzed different aspects of the cosmological frame principle and its possible effects in
cosmological settings. In the first Section, we have introduced a new statement of the principle based on possible stability properties of the conformally-related solutions in the two frames. This statement is useful because it provides a practical way to test the possible physical relevance of a cosmological solution of modified gravity. 

In Sections 2 and 3 we have shown that cosmic
acceleration is a property that respects the frame principle by providing a
dynamical systems analysis of the stability of the solutions of quadratic
gravity with a cosmological constant in both the Jordan and the Einstein
frames. Once one establishes that future acceleration is possible in the
Jordan frame, then the frame principle dictates to also expect it in the
Einstein frame. 

We note that in both frames,  late acceleration is provided by
the cosmological term, cf.  Eqs. (\ref{eq1}) and (\ref{eq2}). As several
authors remark (see for example \cite{agpt,bick}), $f\left(  R\right)  $
gravity models can be viable in different contexts. A characteristic example
is the $R+\epsilon R^{2}$ theory with the $R^{2}$ term producing an
accelerated stage in the early universe preceding the usual radiation and
matter stages \cite{star1,star,mms}. A late-time acceleration in this theory (after the
matter-dominated stage), however, requires a positive cosmological constant in
which case the $R^{2}$ term is no longer responsible for the late-time acceleration.

For $\Lambda<0,$ the potentials (\ref{pote1}) have a negative local minimum,
hence they belong to the class A of the classification in \cite{gmt}. Then,
although flat, these universes recollapse in the Einstein frame according to Theorem 2 in
\cite{gmt}. We conclude that $\Lambda=0$ in (\ref{pote1}) is a bifurcation
value for flat models that recollapse or not. Arguments in  \cite{star1} imply that in the Jordan frame, flat models  eventually recollapse as dust-like models with negative $\Lambda$ in general relativity. A firm belief in the frame principle thus allows us to transfer back to the Jordan frame our earlier result valid in the Einstein frame, and therefore confirm the expectation that in the Jordan frame the Starobinsky result complies with our present calculations.

Our formulation of the frame principle as given in this paper allows for a number of known solutions of modified gravity to be tested for stability in the present context in an effort to decide whether or not they preserve their `physicality' when passing between frames. This  in principle may be applied not only to cosmology but also to other gravitational frameworks, such as black holes or gravity waves, etc. A more elaborate analysis of these problems necessarily involves the further  consideration of structural stability problems associated with the two-conformally-related frames.

\section*{Acknowledgments}
We thank two anonymous referees for useful comments. The  research of SC  was funded by RUDN university,  scientific project number FSSF-2023-0003. JPM thanks the Funda\c c\~ao para a Ci\^encia e Tecnologia (FCT) for the financial support of the grants EXPL/FIS-AST/1368/2021, PTDC/FIS-AST/0054/2021, UIDB/04434/2020, UIDP/04434/2020, and CERN/FIS-PAR/0037/2019.

\end{document}